\documentclass[12pt]{article}
\usepackage{amssymb}
\usepackage{amssymb}
\usepackage{amsmath,amssymb,graphicx}
\usepackage{epsfig}

\evensidemargin=\oddsidemargin 
\newdimen\dummy
\dummy=\oddsidemargin 
\newtheorem{theorem}{Theorem}[section]
\newtheorem{lemma}[theorem]{Lemma}
\newtheorem{proposition}[theorem]{Proposition}
\newtheorem{remark}[theorem]{Remark}

\newcommand{\eps}{\varepsilon}
\newcommand{\hatx}{\widehat{x}}

\newcommand{\calR}{\mathbb{R}}
\newcommand{\nn}{\nonumber}

\newcommand{\be}{\begin{eqnarray}}
\newcommand{\ee}{\end{eqnarray}}
\newcommand{\ben}{\begin{eqnarray*}}
\newcommand{\een}{\end{eqnarray*}}

\begin {document}

\title{\large \bf Homogenization: in Mathematics or Physics?
 }

\author{ Shixin Xu\footnote{Department of Mathematics, University of Science and Technology of China,
Hefei 230026, China (xsxztr@mail.ustc.edu.cn).} \and Changrong
Zhang\footnote{High Speed Aerodynamics Institute, China Aerodynamisc Development and Research Center. Mianyang 622661, China (zcr001@mail.ustc.edu.cn).}
\and Xingye Yue\footnote{Department of Mathematics, Suzhou
University, Suzhou 215006, China (xyyue@suda.edu.cn).} }
\date{}

\maketitle


\begin {abstract}
Homogenization appeared more than 100 years ago. It is an approach
to study the   macro-behavior of a medium by its micro-properties.
In mathematics, homogenization theory considers the limitations of
the sequences of the problems and its solutions when a parameter
tends to zero. This parameter is regarded as the ratio of the
characteristic size in the micro scale to that in the macro scale.
So what is considered is  a sequence of problems in a fixed domain
while the characteristic size in micro scale tends to zero. But for
the real situations in physics or engineering, the micro scale of a
medium is fixed and can not be changed. In the process of
homogenization, it is the size in macro scale which becomes larger
and larger and tends to infinity. We observe that the homogenization
in physics is not equivalent to the homogenization in mathematics up
to some simple rescaling.  With some direct error estimates, we
explain in what means we can accept the homogenized problem as the
limitation of the original real physical problems. As a byproduct, we present some results on the mathematical
homogenization of some problems with source term being only weakly compacted in $H^{-1}$, while in standard homogenization theory,
the source term is assumed to be at least compacted in $H^{-1}$.
A real example is
also given to show the validation of our observation and results.
\end {abstract}


\newpage

\section{Introduction}

Homogenization  appeared more than  100
  years ago. It is an approach to study the
  macro-behavior of a medium by its micro-properties. The origin of
  this word is related to the question of replacement of the
  heterogenous material by an ``equivalent" homogenous one. The earliest
  papers dealing with the problem of this type are \cite{Maxwell-1881}, \cite{Poisson-1822}, and
  the good survey of the results until 1925 is in
 \cite{Lichtenecker-1926}. The name of ``homogenization" was first
  introduced by I. Babuska \cite{Ivo Babuska-1974}.  In
  physics, mechanics and engineering, homogenization is widely used to study the property of
  medium or material by the macro-behavior in stead of the complicated micro structure. The systematic  mathematical theory of
  homogenization was built in \cite{lions-1978}-\cite{Olineak-1994} and so on. But, does the mathematical theory describe
the physics or engineering questions exactly? It seems hard to give a
positive answer. In this paper we take the flow transport problem in
the
   periodic heterogenous porous medium as an example to demonstrate  the difference between the homogenization
    in mathematics and physics. Any other examples such as heat or electric conductivity
    and   mass transfer will lead to the same conclusion.

    In mathematics, we consider the limitation of a sequence $\{u^\epsilon\}_{\epsilon >0}$ such that
\be \label{mathmatics}\left\{
\begin{array}{l}
 -\nabla\cdot (A^{\varepsilon}(x) \nabla u^{\varepsilon}(x)) =f, \ \ \mbox{in}\ \Omega, \\
 \\
u^{\varepsilon}=0, \ \ \ \mbox{on} \ \partial\Omega,
\end{array} \right. \ee
where $\Omega  \in \calR^{d} (d=3)$ is occupied by  the heterogenous
porous medium and the permeability coefficient
$A^{\varepsilon}(x)=A(\frac{x}{\varepsilon})$ with $A(y)$ being
periodic with respect to $y\in Y = [0,1]^d$. $u^{\varepsilon}$ is
the flow pressure in the medium and $f$ is the source.  Just for
simplicity, we take $\Omega=(0,1)^d$. By \cite{lions-1978} and
\cite{Olineak-1994}, under the following assumption: \be(H0): &&f\in
L^2(\Omega)\ and\ there \ exist\ two \
positive\ constants\  \lambda,\ \Lambda\ independent\ of\ \varepsilon,\nn\\
&& such\ that \ \lambda|\xi|^{2}\leq \xi^{T}K(y)\xi\leq
\Lambda|\xi|^{2}, \ \ \forall y, \xi \in \calR^d, \label{uniform}
\ee we have that as $\varepsilon\rightarrow 0$, there exists a
$u^{0}\in H^1_0(\Omega)$ such that $u^{\varepsilon}\rightharpoonup
u^{0}\ weakly\ in\ \ H^{1}_{0}(\Omega)$, where the $u^{0}$ is the solution
of the homogenized problem
 \be \left\{
\begin{array}{l}\label{homo}
 -\nabla\cdot (A^{0} \nabla u^{0}(x)) =f, \ \ \mbox{in}\ \Omega, \\
 \\
u^{0}=0, \  \ \mbox{on} \ \partial\Omega,
\end{array} \right. \ee
with $A^{0}_{ij}=\frac{1}{|Y|}\int_{Y}(A_{ij}+A_{ik}\frac{\partial
N^{j}}{\partial y_{k}})dy$ and  $N^{j}$ being the solution of the
cell problem : \be \left\{
\begin{array}{l}\label{cellp}
 -\nabla_y\cdot (A(y) \nabla_y N^{j}(y)) =\nabla_y\cdot(A(y) e_{j}), \ \ \mbox{in}\ Y = [0,1]^d, \\
N^{j}(y)\ \ \mbox{is Y-periodic and} \ <N^{j}>\triangleq
\frac{1}{|Y|}\int_{Y} N^{j}(y)dy=0.
\end{array} \right. \ee

  Please note that in the above limit process, when the period size $\varepsilon$ tends to zero,
  the domain $\Omega$ does not change, so there will be  more and more $\varepsilon$-periods contained in the whole domain.
  But this is not the case  in the physics. In a physical or engineering problem, the size of the periodic micro
  cell structure can not be changed. What we consider in the physics is  the following problem (take the flow in porous media as an example)
 \be \left\{
\begin{array}{l}\label{physics}
 -\nabla\cdot \left(K\left(\frac{x}{l}\right) \nabla p(x)\right) =f, \ \ \mbox{in}\ \Omega_{D}=(0,D)^{d}, \\
 \\
p=0, \ \ \ \mbox{on} \ \partial\Omega_{D},
\end{array} \right. \ee
where $l (m)$ and $D (m) $ are the characteristic lengthes of the
periodic micro cell and the whole medium respectively and $l\ll D$, $p\ (N/m^2 =
Kg /m/ s^2)$ is the pressure, $K\ (m^3 s/kg)$ is the permeability
coefficient and the source term $ f\ (\frac 1 s) = \frac {f_0}\rho$
with $f_0 \ (kg/m^3/s)$ and the density $\rho\ (kg/m^3)$.

We want to get the effective coefficient and the homogenized problem
in physics and try to explain in what sense  we can expect this
result is valid. In the early work of I. Babuska \cite{babuska-1975}
on the homogenization approach in engineering, he first pointed out
that: `$l$ is a given parameter, with physical meaning which cannot
be changed, e.g. cannot be made ``sufficiently'' small'. In his
other work \cite{babuska-1976}, \cite{babuska-1977}, he mentioned
that
 $\varepsilon$ in \eqref{mathmatics} is the ratio of
  micro-scale (cell scale) to the macro-scale. If we set $\eps =
\frac l D$ and $l$ is fixed, then  $\varepsilon\rightarrow 0$ means that
  $D$ tends to infinity. The problem is if we take the transformation $\eps =
\frac l D$, can we transfer the physical  problem \eqref{physics} to
the mathematical problem \eqref{mathmatics} while the Assumption
(H0) is still valid?

Let us take a variable transformation $\hatx = \frac x D$ {\em
mathematically}, then the pressure $p(\hatx)$ satisfies that \be
\label{math-model} \left\{
\begin{array}{l}
 - \nabla_{\hatx} \cdot\left( D^{-2}K\left(\frac{\widehat{x}}{\eps}\right) \nabla_{\hatx} p(\widehat{x})\right) = f(\hatx), \ \ \mbox{in}\ \Omega=(0,1)^{d}, \\
 \\
p(\hatx)=0.\ \ \ \mbox{on}\ \partial\Omega.
\end{array} \right. \ee
It is obvious that   the Assumption (H0) can  not be satisfied even
for a simple  example as $f\equiv 1$, since $
D^{-2}K(\frac{\widehat{x}}{\eps})\rightarrow 0$, as $\eps
\rightarrow 0$.

 Let's go back to {\em physics}. What's the meaning of the variable
transformation  $\hatx = \frac x D$ from \eqref{physics} to
\eqref{math-model}?  The only difference is the {\em unit} of
length: in \eqref{physics}, the length unit is one meter; in
\eqref{math-model}, the length unit is $D$ meters.

Therefore, if we take a variable transformation $\hatx = \frac x D$
{\em physically}, then the pressure $\widehat{p}(\hatx)$ should
satisfy the same Conservation Law of Mass, which is independent of
the unit in length,
 \be \left\{
\begin{array}{l}
 -\widehat{\nabla}\cdot \left(\widehat{K}\left(\frac{\widehat{x}}{\eps}\right) \widehat{\nabla} \widehat{p}(\widehat{x})\right) =\widehat{f}, \ \ \mbox{in}\ \Omega=(0,1)^{d}, \\
 \\
\widehat{p}=0, \ \ \ \mbox{on}\ \partial\Omega,
\end{array} \right. \label{phy-model} \ee
but all the physical quantities: the pressure $\widehat{p}$, the
permeability coefficient $\widehat{K}$, the source term
$\widehat{f}$ and the gradient operator $\widehat{\nabla}$ are all
measured in the new length scale, i.e.
$$
\widehat{x}=\frac{x}{D}\ (Dm), \ \ \widehat{K}=\frac{K}{D^{3}}\
((Dm)^3s/kg), \ \ \widehat{p}=Dp\ (kg/Dm/s^2), $$
$$\widehat{\nabla}=D\nabla\ ( 1/{Dm}), \ \ \widehat{f}=f\ ( 1/s).$$

It should be noted that $\widehat{K}=\frac{K}{D^{3}}$ will tend to
zero as D tending to infinity, since $K$ is the fixed physical
quantity in the original unit system. So    the Assumption ({H0})
can not be valid either and  we can not expect the validity of the
 homogenization theory even for  a constant source term.

So far, we see that the homogenization of the physical problem
\eqref{physics}
 can not easily fall into the mathematical framework \eqref{mathmatics}-\eqref{cellp}
 by a direct transformation in mathematics or  physics.
  In the following,  we assume that $D$ is sufficient large but fixed.
  We consider two different situations for the source term $f$. The first
  situation is that $f$ has no micro-structure, and the other
  is that $f$ has micro structure with period $(0,l)^d$, for example, the source term may have the form as $f = f_1(x,x/l)+\nabla \cdot f_2(x,x/l)$.
  We first give
the homogenized  problem with the effective coefficient $K^{0}$ for
different situations. We  present the error estimate between the
pressure $p$ and the first order expansion $p_1$, then try to
understand in what sense the homogenized problem is a limitation of
the original problem. It is worthwhile to point out that  so far the
 mathematical homogenization theory (for which we consider the limitation
  as $l\rightarrow 0$) is still incomplete for the second situation, since
  the source term is only convergent weakly in $H^{-1}$(see \cite{Doina Cioranescu and
Patrizia Donato} and \cite{taoyu}). The homogenization theory for
this kind of problem may have independent interests.

The outline of paper is as follows: in $\S2$ we discuss the
situation that the source term has no micro-structure; in $\S3$ we
discuss the situation that the source term has micro-structure; in
$\S4$ a real example is given to show the validation of our
observation and results.

\section {The source term  has no micro-structure }
\setcounter{equation}{0}

we say $f$ has no micro-structure if $f$ does not contain any
micro-scale information
 at the scale comparable to or less than   $l$.  We discuss two different cases in this situation.
\begin{itemize}
\item Case a: $f\in L^2(R^d)$;\ \ \ \  \ Case b: $f\in L^{\infty}(R^d)$.
\end{itemize}

By unit transformation, we get
 (\ref{phy-model}), which we have known do not satisfy Assumption (H0).
In order to let  the
 Assumption (H0) be valid, we introduce a new setting of problem as
  \be \left\{
\begin{array}{l}\label{physicsmath}
 -\widehat{\nabla}\cdot \left(K^{\varepsilon}(\widehat{x}) \widehat{\nabla} p^{\varepsilon}(\widehat{x})\right) =\overline{f}(\hatx), \ \ \mbox{in} \ \Omega=(0,1)^{d}, \\
 \\
p^{\varepsilon}=0, \ \ \ \mbox{on} \ \partial\Omega,
\end{array} \right. \ee
 with  \begin{eqnarray}\label{measurealt1}\textbf{a:}\left\{
\begin{array}{l}
 K^{\varepsilon}(\widehat{x})=K(\frac {\hatx}{\eps})=D^{3}\widehat{K}(\frac{\widehat{x}}{\varepsilon}),\\
 p^{\varepsilon}(\widehat{x})=D^{-\frac 3
 2}\widehat{p}(\widehat{x}),\\
 \overline{f}(\hatx)=D^{\frac 3 2}f(\hatx),
\end{array} \right.
\textbf{b:}\left\{
\begin{array}{l}
 K^{\varepsilon}(\widehat{x})=K(\frac {\hatx}{\eps})=D^{3}\widehat{K}(\frac{\widehat{x}}{\varepsilon}),\\
 p^{\varepsilon}(\widehat{x})=D^{- 3}\widehat{p}(\widehat{x}),\\
 \overline{f}(\hatx)=f(\hatx).
\end{array} \right.\end{eqnarray}
%
%
It is easy to check that Assumption (H0) is satisfied for the both
 cases.


 Then we know from homogenization theory (see \cite{V.V.Zhikov-2006}) that
 there exists a $\widetilde{p_0}$ such that
 \be \left\{
\begin{array}{l}\label{convergencemath}
p^{\eps}\rightharpoonup \widetilde{p_0} \ \mbox{weakly in}\
H^{1}_{0}(\Omega),\\
\\
 K^{\eps}\widehat{\nabla}p^{\eps}\rightharpoonup\widetilde{K^{0}}\widehat{\nabla}\widetilde{p_0}\ \mbox{weakly in}
 \ (L^{2}(\Omega))^{d},\\
\end{array} \right. \ee
where $\widetilde{p_0}$ is the  solution of  the
homogenized problem of (\ref{physicsmath})  :
 \be \left\{
\begin{array}{l}\label{mathhomo}
 -\widehat{\nabla}\cdot \left(\widetilde{K^{0}}\widehat{\nabla} \widetilde{p_0}(\widehat{x})\right) =\overline{f}, \ \ \mbox{in} \ \Omega=(0,1)^{d}, \\
 \\
\widetilde{p_0}=0, \ \ \ \mbox{on} \ \partial\Omega,
\end{array} \right. \ee
with
$\widetilde{K_{ij}^{0}}=\int_{Y}(K_{ij}(y)+K_{ik}(y)\frac
{\partial \widetilde{N^{j}}(y)}{\partial y_{k}})dy$
 and $\widetilde{N}^{j}(\frac {\hatx}{\eps})=\widetilde{N}^{j}(y)$ solving of the cell problem: \be \left\{
\begin{array}{l}\label{mathcell}
 -\widehat{\nabla}_y\cdot (K(y) \widehat{\nabla}_y(\widetilde{N}^{j}(y)+y_{j}) )=0, \ \ \mbox{in}\ Y=(0,1)^{d}, \\
 \\
\widetilde{N}^{j}\ \mbox{is Y-periodic},
\int_Y\widetilde{N}^{j}(y)dy =0.
\end{array} \right. \ee

If we denote by
\begin{equation}\label{mafirstoder}
\widetilde{p_1}=\widetilde{p_0}+\varepsilon \widetilde{N}^{j}
\frac{\widehat{\partial}\widetilde{p_0}}{\widehat{\partial}\widehat{x}_{j}},
\end{equation}then there exits a positive constant  C independent of
$\eps$ such that
\begin{equation}\label{mathestimate}
\|p^{\varepsilon}-\widetilde{p_1}\|_{H^{1}(\Omega)}\leq C
\varepsilon^{\frac{1}{2}}.
\end{equation}
After carefully checking the exact dependence of the constant C in
(\ref{mathestimate}), we have

\begin{proposition}\label{theorem for estimate}
If the coefficient and the source term of equation
(\ref{physicsmath})
 satisfy the Assumption (H0), then there exists a positive constant C
 independent of $\eps,\ p^{\eps},\ \widetilde{p_0}$, such that
 \begin{eqnarray}\label{fineestimate}
 \|\widehat{\nabla}(p^{\varepsilon}-\widetilde{p_1})\|_{L^{2}(\Omega)}
&\leq&C\varepsilon\frac{1}{\lambda^2}(\|G\|_{L^{\infty}(Y)}+\Lambda\|N\|_{L^{\infty}(Y)})\|\overline{f}(\hatx)\|_{L^2(\Omega)}\nonumber\\
& &+C\varepsilon^{\frac{1}{2}}(\frac
{\Lambda}{\lambda^2})(1+\|N\|_{L^{\infty}(Y)})\|\overline{f}(\hatx)\|_{L^2(\Omega)},
 \end{eqnarray}
\begin{equation}\label{l2estimate}
\|p^{\varepsilon}-\widetilde{p_0}\|_{L^{2}(\Omega)}\leq C\varepsilon\frac 1
{\lambda^2}(\|G\|_{L^{\infty}(Y)}+\Lambda\|N\|_{L^{\infty}(Y)})\|\overline{f}(\hatx)\|_{L^2(\Omega)}.
\end{equation}
Here $G=(G^1,...,G^d)$ with $G^j$ being a skew-symmetrical matrix
satisfying (see \cite{Olineak-1994})\be \frac {\partial}{\partial
y_k}G^j_{ik}=K(y)_{ij}+K(y)_{ik}\frac{\partial N^j}{\partial
y_k}-\widetilde{K}^0_{ij},\ j=1,...,d.\ee
As to energy, we have
 \be
|\int_{\Omega}\widehat{\nabla}p^{\eps}\cdot
K^{\eps}\widehat{\nabla}p^{\eps}-\int_{\Omega}\widehat{\nabla}\widetilde{p_0}\cdot\widetilde{K^0}\widehat{\nabla}\widetilde{p_0}|
\rightarrow 0,\ \mbox{as}\ \eps\rightarrow 0.\ee
\end{proposition}


%

By the inverse transformation of (\ref{measurealt1}) and changing
the unit in length  from $Dm$ to $m$, we  obtain the following {\em homogenized}  problem of \eqref{physics} from (\ref{mathhomo}) and
(\ref{mathcell}) 
 \be\label{phyhomo} \left\{
\begin{array}{l}
 -\nabla\cdot (K^{0} \nabla p_0(x)) =f, \ \ \mbox{in}\ \Omega_{D}=(0,D)^{d}, \\
\\
p_0=0, \ \ \ \mbox{on}\ \partial\Omega_{D},
\end{array} \right. \ee
with $K^{0}$  determined as follows:
\begin{eqnarray}\label{homomatrix}
K_{i,j}^{0}&=&\int_{Y}\left(K_{ij}(y)+K_{ik}(y)\frac
{\partial N^{j}(y)}{\partial y_{k}}\right)dy\nonumber\\
&=&\frac {1} {| l
|^{d}}\int_{(0,l)^{d}}\left(K_{ij}\left(\frac{x}{l}\right)+lK_{ik}\left(\frac {x}{l}\right)\frac
{\partial N^{j}\left(\frac {x}{l}\right)}{\partial x_{k}}\right)dx,
\end{eqnarray}
where ${\displaystyle N^{j}\left(\frac {x}{l}\right)=N^j(y)\ (j=1...d)}$ is the solution of cell
problem \be \left\{
\begin{array}{l}\label{phycelly}
 -\nabla_y\cdot (K(y) \nabla_y(N^{j}(y)+y_{j}) )=0, \ \mbox{in} \ Y=(0,1)^{d}, \\
\\
N^{j}\ \mbox{is Y-periodic and}\ \int_{Y}{N}^{j}(y)dy =0.
\end{array} \right. \ee
It's easy to find that \eqref{phycelly} is equivalent to the
following problem \be \left\{
\begin{array}{l}\label{phycell}
 -\nabla\cdot (K(\frac x l) \nabla (lN^{j}(\frac x l)+x_{j}) )=0, \ \mbox{in}\ Y_l=(0,l)^{d}, \\
\\
N^{j}\ \mbox{is}\ Y_l\ \mbox{periodic and}\ \frac{1}{l^d}\int_{(0,l)^d}{N}^{j}(\frac x l)dx
=0.
\end{array} \right. \ee
\begin{remark}
In fact, if we simply regard the micro size $l$ in \eqref{physics} as a small parameter and formally apply the mathematical homogenization theory, we would obtain the same homogenization settings for
\eqref{physics} as \eqref{phyhomo}-\eqref{phycell}.  Furthermore we would still have the following mass balance or homogenization rule as :
for any $\eta\in R^d$, if $p_{\eta}$ solves \be\left\{
\begin{array}{l}\label{remarkequantion}
-\nabla\cdot (K(\frac x l) \nabla p_{\eta} )=0, \ \mbox{in}\ Y_l=[0,l]^{d}, \\
\\
p_{\eta}-\eta\cdot x\ \mbox{is periodic in}\
[0,l]^d,\end{array}\right. \ee then $\langle \nabla
p_{\eta}\rangle=\frac 1 {l^d}\int_{Y_l}\nabla p_{\eta}dx=\eta$ and
by  (\ref{homomatrix}) and (\ref{phycell}) ,we have
\begin{equation}\langle K(\frac x l)\nabla p_{\eta}\rangle_{R^d}=\langle
K(\frac x l)\nabla p_{\eta}\rangle_{Y_l}=K^0\langle \nabla
p_{\eta}\rangle_{Y_l}=K^0\langle \nabla
p_{\eta}\rangle_{R^d},\end{equation} which means the mass balance
between the micro and macro scales. This is the reason why the homogenized coefficient $K^0$ is
also called as the effective coefficient of $K$.
\end{remark}

 The relationship between $p_0(x)$ and
$\widetilde{p_0}(\hatx)$ are
\begin{eqnarray}\label{relationship1}\textbf{a:}\left\{
\begin{array}{l}
 p_0(x)=
 D^{-\frac 1 2}\widetilde{p_0}(\hatx),\\
 \\
 K^0=\widetilde{K^0},
\end{array} \right.
\textbf{b:}\left\{ \begin{array}{l}
 p_0(x)=
 D^2\widetilde{p_0}(\hatx),\\
 \\
 K^0=\widetilde{K^0}.
\end{array} \right.
\end{eqnarray}


By (\ref{relationship1}) and Proposition (\ref{theorem for estimate}), we can obtain the next theorem
\begin{theorem}If $p$ is the solution of \eqref{physics}, $p_1$ is defined as follows \begin{equation}\label{phyfirstod}
p_1=p_0+l N^{j} \frac{\partial p_0}{\partial x_{j}}.
\end{equation} and $f\in L^2(R^3)$, then
there exists a positive constant C
independent of D such that
\be\label{final physics estimate1a for grad}
\int_{\Omega_D}|\nabla(\frac {p} {D}-\frac {p_1}{D})|^2dx &\leq&
C(\frac l
D)^2\|f(x)\|^2_{L^2(\Omega_D)}+C\frac{l}{D}\|f(x)\|^2_{L^2(\Omega_D)}.
\ee

\begin{equation}\label{l2estimate1a} \int_{\Omega_D}|\frac p
{D^2}-\frac {p_0} {D^2}|^2dx\leq C(\frac l
D)^2\|f(x)\|^2_{L^2(\Omega_D)}.
\end{equation}
As to energy, there exists a positive const C independent of D, such that
 \be |\frac 1
{D^3} \int_{\Omega_D} \nabla p\cdot K\nabla p-\frac 1 {D^3}
\int_{\Omega_D} \nabla p_0\cdot K^0\nabla p_0|\leq C\frac 1 D,\ee
which means the convergence of the density of energy;
\end{theorem}

\begin{theorem}
If $p$ is the solution of \eqref{physics}, $p_1$ is defined in \eqref{phyfirstod} and $f\in L^{\infty}(R^3)$, then there exists a positive const C independent of D such that

\begin{eqnarray}\label{final physics estimate1b for grad}
\frac 1 {D^3}\int_{\Omega_D}|\nabla(\frac {p} {D}-\frac
{p_1}{D})|^2dx &\leq& C(\frac l
D)^2\|f(x)\|^2_{L^{\infty}(\Omega_D)}+C\frac{l}{D}\|f(x)\|^2_{L^{\infty}(\Omega_D)}.\end{eqnarray}
\be\label{l2estimate1b} \frac 1 {D^3} \int_{\Omega_D}|\frac p
{D^2}-\frac {p_0} {D^2}|^2dx \leq C(\frac l
{D})^2\|f(x)\|^2_{L^{\infty}(\Omega_D)}.\ee
As to energy, we  have
\be |\frac 1 {D^3}\int_{\Omega_D}\nabla\frac p D\cdot
K\nabla\frac p D dx-\frac 1 {D^3}\int_{\Omega_D}\nabla\frac {p_0}
D\cdot K^0\nabla\frac {p_0} D dx|\rightarrow 0, \ \mbox{as
D}\rightarrow \infty.\ee
\end{theorem}
The above theorems explain in what sense we can accept the homogenized problem \eqref{phyhomo}.
%
%
\section {The source term  has micro-structure }
\setcounter{equation}{0}
In this situation, we will discuss the source term with the
following  micro-structure,
 \be
\left\{
\begin{array}{l}\label{physics for situation 2}
 -\nabla\cdot \left(K(\frac{x}{l}) \nabla p(x)\right) =f\left(x, \frac x l\right)+\nabla\cdot F\left(x, \frac x l\right), \ \ \mbox{in}\ \Omega_{D}=(0,D)^{d}, \\
 \\
p=0, \ \ \ \mbox{on} \ \partial\Omega_{D}.
\end{array} \right. \ee
By unit transformation, we get
 \be \left\{
\begin{array}{l}
 -\widehat{\nabla}\cdot \left(\widehat{K}\left(\frac{\widehat{x}}{\eps}\right) \widehat{\nabla} \widehat{p}(\widehat{x})\right) =f\left(\hatx, \frac {\hatx}{\eps}\right)+\widehat{\nabla}\cdot \widehat{F}\left(\hatx, \frac {\hatx} {\eps}\right), \ \ \mbox{in}\ \Omega=(0,1)^{d},
\\
 \\
\widehat{p}=0, \ \ \ \mbox{on}\ \partial\Omega.
\end{array} \right. \ee
Setting \be \left\{
\begin{array}{l}\label{measurealt2}
 K^{\varepsilon}(\widehat{x})=D^{3}\widehat{K}(\frac{\widehat{x}}{\varepsilon}),\\
 p^{\varepsilon}(\widehat{x})=D^{- 3}\widehat{p}(\widehat{x}),\\
\end{array} \right.\ee
we obtain
 \be \left\{
\begin{array}{l}\label{physicsmath for situation2}
 -\widehat{\nabla}\cdot \left(K^{\varepsilon}\left(\widehat{x}\right) \widehat{\nabla} p^{\varepsilon}(\widehat{x})\right) =f\left(\hatx,\frac {\hatx}{\eps}\right)+\widehat{\nabla}\cdot \widehat{F}\left(\hatx,\frac {\hatx} {\eps}\right), \ \ \mbox{in} \ \Omega=(0,1)^{d}, \\
 \\
p^{\varepsilon}=0, \ \ \ \mbox{on} \ \partial\Omega.
\end{array} \right. \ee
The homogenization  for this kind of problem may have independent
interest, since the source term here is only weakly convergent in
$H^{-1}(\Omega)$ as $\eps\rightarrow 0$. The standard theory only
treats the case that the source term is strongly convergent in
$H^{-1}(\Omega)$ see(\cite{Olineak-1994}).In \cite{Doina Cioranescu
and Patrizia Donato} and \cite{taoyu}   some incomplete
results were present for this case. We will
 establish the homogenization theory for \eqref{physicsmath for situation2}.

In the beginning, we  introduce an important lemma that will be used
later.
\begin{lemma}\label{f H-1 estimate}
 If $f(x,y)\in L^{\infty}(Y,C^{0,1}(\Omega))$ and is Y-period with respect to y, where $\Omega$ is an arbitrary bounded open subset of
 $R^d$ and $Y=[0,1]^d$,
 then $f(x,\frac x {\eps})\rightarrow Mf(x)\triangleq \frac{1}{|Y|}\int_Yf(x,y)dy$ in $H^{-1}(\Omega)$ and there exists a constant
 $C\geq0$ such that
 \begin{equation}
 \|f(x, \frac x {\eps})-Mf(x)\|_{H^{-1}(\Omega)}\leq C\eps\|f\|_{ L^{\infty}(Y,\ C^{0,1}(\Omega))},
 \end{equation}
\end{lemma}
The proof of lemma  is similar to the lemma 1.6 in \cite{Olineak1992}.
\begin{theorem}\label{lemma on  homgenzition for situation 2 }
If $u^{\eps}$ is the solution of the following
problem\be\left\{\begin{array}{l}\label{lemma problem}
-\nabla\cdot(A^{\eps}(x)\nabla u^{\eps})=f\left(x, \frac x {\eps}\right)+\nabla\cdot F\left(x, \frac x {\eps}\right),\ \mbox{in}\ \Omega,\\
\\
u^{\eps}=0,\ \mbox{on}\ \partial\Omega,\end{array}\right.\ee where
$A^{\eps}(x)=A\left(\frac x {\eps}\right)$ is symmetric satisfying uniformly elliptic condition, i.e. there exist
two positive constants $\lambda$, $\Lambda$ independent of
$\varepsilon$, such that $ \lambda|\xi|^{2}\leq \xi^{T}K(y)\xi\leq
\Lambda|\xi|^{2}, \ \forall y, \xi \in \calR^d$, and A(y) is Y-period.
$f(x,y)$ and $F(x,y)$ are bounded and  Y-period with respect to y
 then as $\eps\rightarrow 0$, \be\label{convergence in lemma}\left\{\begin{array}{l}
u^{\eps}\rightharpoonup u^0 \ \mbox{weakly in}\
H^1_0(\Omega),\\
A^{\eps}\nabla u^{\eps}+F(x,\frac x {\eps})\rightharpoonup A^0\nabla
u^0+F^0,\ \mbox{weakly in}\ (L^2(\Omega))^d.\end{array}\right.\ee
$u^0$ is the solution of the following homogenized problem
\be\left\{\begin{array}{l}\label{lemma homogenized pr}
-\nabla\cdot(A^{0}\nabla u^{0})=f^{0}(x)+\nabla \cdot F^{0}(x),\ \mbox{in}\ \Omega,\\
\\
u^{0}=0,\ \mbox{on}\ \partial\Omega,\end{array}\right.\ee with
$A^0$, $f^0$, and $F^0$
 defined as follows
 \be\label{lemma homomatrix}\left\{\begin{array}{l}
 A_{ij}^0=\langle A_{ij}(y)+A_{ik}\frac{\partial N^j}{\partial
y_k}\rangle_Y,\\
f^0(x)=\langle f(x, y)\rangle_Y,\\
F^0(x)=\langle F(x,
y)+A(y)\nabla_yw(y)\rangle_Y,\end{array}\right.\ee and $N^j(y)$,
$w(x,y)$ solving the cell problems : \be\label{lemma cell
N}\left\{\begin{array}{l}
-\nabla_y\cdot(A(y)\nabla_y N^j(y))=\nabla_y\cdot(A(y)e_j),\ \mbox{in}\ Y,\\
\\
N^j(y)\ \mbox{is periodic in Y}, \langle N^j\rangle_Y=0,
\end{array}\right.\ee
\be\left\{\begin{array}{l}\label{lemma cell w}
-\nabla_y\cdot(A(y)\nabla_y w(x,y))=\nabla_y\cdot(F(x, y)),\ \mbox{in}\ Y,\\
\\
w(x,y)\ \mbox{is periodic in Y}, \langle w\rangle_Y=0.
\end{array}\right.\ee

Further more, if we denote by \begin{equation}\label{lemma p1} u_1=u^0+\eps
N^j\frac{\partial u^0}{\partial x_j}+\eps w,
\end{equation} and  assume that $A(y),\ f(x,y),\ F(x,y)$ are smooth enough and $w(x,y)\in W^{1,\infty}(\Omega\times Y)$,
then there exists a positive constant C independent of $\eps$ such that
\be\label{estimate for situation 2}\|\nabla(u^{\eps}-u_1)\|_{L^2(\Omega)}\leq C\eps^{\frac 1 2}\ee
\end{theorem}
\textbf{Proof.} By the standard asymptotic expansion method, we can
get the equations (\ref{lemma homogenized pr})-(\ref{lemma p1}) (\cite{Doina Cioranescu and  Patrizia Donato}, \cite{taoyu}).
  We  first use
Tartar's method (\cite{Doina Cioranescu and  Patrizia Donato}) to prove
(\ref{convergence in lemma}). We denote by $f^{\eps}=f(x,\frac x {\eps})$ and $F^{\eps}=F(x,\frac x {\eps})$ for short.
By the regularity of elliptic equation,  we obtain $u^{\eps}$ is bounded in $H^1_0(\Omega)$ and $\xi^{\eps}$ is bounded in $(L^2(\Omega))^d$, where $\xi^{\eps}=A^{\eps}\nabla u^{\eps}+F^{\eps}$ is a vector-function and satisfies
\be\label{xiepseq}\int_{\Omega}\xi^{\eps}\nabla v dx=\langle f^{\eps}, v\rangle_{H^{-1}(\Omega),H^1_0(\Omega)},\ \ \forall\ v\in\ H^1_0(\Omega).\ee
By the compact property,  there exists a subsequence (still denoted by $\eps$), such that
\be\left\{\begin{array}{l}
u^{\eps}\rightharpoonup u^0\ \mbox{weakly in}\ H^1_0(\Omega),\\
\xi^{\eps}\rightharpoonup \xi^0\mbox{weakly in}\ (L^2(\Omega))^d, \end{array}\right.\ee
Taking $\eps\rightarrow 0$ in \eqref{xiepseq}, we have
\be\label{xi0eq}\int_{\Omega}\xi^{0}\nabla v dx=\langle f^{0}, v\rangle_{H^{-1}(\Omega),H^1_0(\Omega)},\ \ \forall\ v\in\ H^1_0(\Omega).\ee
Therefor, (\ref{convergence in lemma}) is proved if we show that \be \xi^0=A^0\nabla u^0+F^0.\ee
If we set \be\chi_{\lambda}^{\eps}=\lambda\cdot x+\eps N_{\lambda}\left(\frac x {\eps}\right),\ee with $\lambda\in R^d$ and $N_{\lambda}(y)$ solving the following problem\be
\left\{\begin{array}{l}\label{Nlambda}
-\nabla_y\cdot(A(y)\nabla_y N_{\lambda}(y))=\nabla_y\cdot(A(y)\lambda),\ \mbox{in}\ Y,\\
\\
N_{\lambda}\ \mbox{is periodic in }\ Y\  and\  \langle N_{\lambda}\rangle_Y=0,
\end{array}\right.\ee
the we have the following limitation:\be \left\{\begin{array}{l}\label{chiconvergence}\ \chi_{\lambda}^{\eps}\rightharpoonup \lambda\cdot x,\ \mbox{weekly in} H^1(\Omega),\\
\ \chi_{\lambda}^{\eps}\rightarrow \lambda\cdot x,\ \mbox{strongly in} L^2(\Omega).\end{array}\right.\ee
Introduce the vector function \be \eta_{\lambda}^{\eps}=A^{T}\nabla_y\chi_{\lambda}^{\eps}.\ee
By the definition of $\chi_{\lambda}^{\eps}$ and \eqref{Nlambda}, we can easily obtain
\be\left\{\begin{array}{l}\label{etaeq}
\eta_{\lambda}^{\eps}\rightharpoonup (A^0)^T\lambda \ \mbox{weakly in }\ L^2((\Omega))^d,\\
\int_{\Omega}\eta_{\lambda}^{\eps}\cdot \nabla v dx=0,\ \forall\ v\in\ H^1_0(\Omega).\end{array}\right.\ee
For any $\varphi\in \mathcal{D}(\Omega)$,  choose $\varphi\chi_{\lambda}^{\eps}$ as the test function in \eqref{xiepseq} and $\varphi u^{\eps}$ as the test function  in \eqref{etaeq}. We have
\be\label{xieq2} \int_{\Omega}\xi^{\eps}\cdot \nabla\chi_{\lambda}^{\eps}\varphi dx+\int_{\Omega}\xi^{\eps}\cdot\nabla\varphi\chi_{\lambda}^{\eps}dx=\langle f^{\eps}, \varphi\chi_{\lambda}^{\eps}\rangle_{H^{-1}(\Omega),H^1_0(\Omega)},\ \forall\ \varphi\in \mathcal{D}(\Omega), \ee
\be\label{etaqu2} \int_{\Omega}\eta_{\lambda}^{\eps}\cdot \nabla u^{\eps}\varphi dx+\int_{\Omega}\eta_{\lambda}^{\eps}\cdot\nabla\varphi u^{\eps}dx=0,\ \forall\ \varphi\in \mathcal{D}(\Omega). \ee
By the definition of $\eta_{\lambda}^{\eps}$, we have
$$A^{\eps}\nabla u^{\eps}\cdot\nabla\chi_{\lambda}^{\eps}=(A^{\eps})^T\nabla\chi_{\lambda}^{\eps}\cdot \nabla u^{\eps}=\eta_{\lambda}^{\eps}\cdot\nabla u^{\eps}. $$
From \eqref{xieq2}-\eqref{etaqu2}, we have
\be \int_{\Omega}\xi^{\eps}\cdot\nabla\varphi\chi_{\lambda}^{\eps}dx+\int_{\Omega}F^{\eps}\cdot\nabla\chi_{\lambda}^{\eps}\varphi dx-\int_{\Omega}\eta_{\lambda}^{\eps}\cdot\nabla\varphi u^{\eps}dx=\langle f^{\eps}, \varphi\chi_{\lambda}^{\eps}\rangle_{H^{-1}(\Omega),H^1_0(\Omega)}.\ee
Taking $\eps \rightarrow 0$, by \eqref{chiconvergence} and \eqref{etaeq}, we obtain
\begin{eqnarray}
&&\int_{\Omega}\xi^{0}\cdot\nabla\varphi(\lambda\cdot x)dx+\int_{\Omega}\varphi dx\int_{Y}F(x,y)\cdot(\lambda+\nabla_yN_{\lambda})dy-\int_{\Omega}\lambda (A^0)^T\cdot\nabla\varphi u^0 dx\nonumber\\&& \ \hspace{1cm}=\langle f^{0}, (\lambda\cdot x)\varphi\rangle_{H^{-1}(\Omega),H^1_0(\Omega)},\end{eqnarray}
which can be rewritten  in the form\begin{eqnarray}
&&\int_{\Omega}\xi^{0}\cdot\nabla[\varphi(\lambda\cdot x)]dx-\int_{\Omega}\xi^{0}\cdot\varphi\lambda dx+\int_{\Omega}\varphi dx\int_{Y}F(x,y)\cdot(\lambda+\nabla_yN_{\lambda})dy\nonumber\\ && \ \hspace{1cm}-\int_{\Omega}\lambda (A^0)^T\cdot\nabla\varphi u^0 dx=\langle f^{0}, (\lambda\cdot x)\varphi\rangle_{H^{-1}(\Omega),H^1_0(\Omega)},\end{eqnarray}
By \eqref{xi0eq}, we get
\be\int_{\Omega}\xi^{0}\cdot\varphi\lambda dx=\int_{\Omega}\varphi dx\int_{Y}F(x,y)\cdot(\lambda+\nabla_yN_{\lambda})dy+\int_{\Omega}\lambda (A^0)^T\cdot\nabla u^0\varphi  dx.\ee
If let $\lambda=e_i$, we can obtain \be\xi^{0}_i=(A^0\nabla u^0)_i+F^0_i.\ee
Here we use the following relationship (\cite{taoyu})
\be F_i^0=\int_Y(F_i(x,y)+(A(y)\nabla_y w)_i)dy=\int_Y(F(x,y)(e_i+\nabla_y N^i))dy.\ee
Then we have proved \eqref{convergence in lemma}.

Next we will give the error estimate \be\|\nabla(u^{\eps}-u_1)\|_{L^2(\Omega)}\leq C\eps^{\frac 1 2}.\ee Following the argument in \cite{Olineak-1994},  we first compute
\begin{eqnarray}
A^{\eps}\nabla u_1-A^{0}\nabla u^{0}& = &
\left(A^{\eps}\left(e_{j}+\nabla_{y}N^{j}\right)-A^{0}e_{j}\right)\frac{\partial
u^{0}}{\partial x_{j}}+\eps A^{\eps}N^{j}\nabla(\frac{\partial u^{0}}{\partial x_j})\nonumber\\
&=&g^{j}(y)\frac{\partial u^{0}}{\partial x_{j}}+\varepsilon
A^{\varepsilon}N^{j}\nabla(\frac{\partial u^{0}}{\partial x_j}),\nonumber\\
\end{eqnarray}
with $g^{j}(y)=A^{\varepsilon}(e_{j}+\nabla_{y}N^{j})-A^{0}e_{j}.$
By the cell problem (\ref{lemma cell N}) and the definition of
$A^{0}$ in (\ref{lemma homomatrix}), we find \begin{equation} \langle
g^{j}(y)\rangle_Y=0\ \ and\ \ \nabla_y\cdot g^{j}(y)=0 .\end{equation}
Then there exists a skew-symmetric matrix $G^{j}$ (\cite{Olineak-1994}), such that
\begin{equation}\label{gproperty} g^{j}=\nabla_y\cdot
G^{j},\ \ G^{j}_{ik}=-G^{j}_{ki},\ \ and \ \ \ G^{j}_{ik}\in
H^{1}_{per}(Y).
\end{equation}
By this property, we can get:
\begin{equation}
g^{j}\frac{\partial u^{0}}{\partial
x_{j}}=\varepsilon\nabla\cdot\left(\frac{\partial u^{0}}{\partial
x_{j}}G^{j}\right)-\varepsilon G^{j}\nabla\left(\frac{\partial
u^{0}}{\partial x_{j}}\right),
\end{equation}
and
\begin{eqnarray}\label{error equation for p 1}
-\nabla\cdot (A^{\varepsilon} \nabla (u^{\varepsilon}-u_{1}))
&=&R^{\eps}+\nabla\cdot r_{\eps},
\end{eqnarray}
with $R^{\eps}=\nabla\cdot\left( F(x,y)+A(y)\nabla_yw(x,y)-
F^0(x)\right) +(f(x,y)-f^0(x))$, and $r^{\eps}=\eps
G_{ik}^j\frac{\partial^2u^0}{\partial x_j\partial x_k}+\eps
A^{\eps}_{ij}N^j\frac{\partial^2u^0}{\partial x_j\partial x_k}.$

If we introduce the boundary corrector $\theta^{\eps}$ as \be\left\{\begin{array}{l}\label{theta equation}
-\nabla\cdot (A^{\varepsilon}(x) \nabla\theta^{\eps})=0,\
\mbox{in}\ \Omega,\\
\\
\theta^{\eps}=-\eps N^j\frac{\partial u_0}{\partial x_j}-\eps w,\
\mbox{on}\ \partial\Omega,
\end{array}\right.\ee
then $e_{\eps}=u^{\varepsilon}-u_{1}-\theta^{\eps}\in H^1_0(\Omega)$
satisfies
\begin{eqnarray}\label{error equation for p 12}
-\nabla\cdot (A^{\varepsilon} \nabla
(u^{\varepsilon}-u_{1}-\theta^{\eps})) &=&R^{\eps}+\nabla\cdot
r_{\eps},
\end{eqnarray}and the weak form is
\begin{equation}
\int_{\Omega}A^{\eps}\nabla e_{\eps}\nabla \phi dx=\int_{\Omega}R^{\eps}\phi dx+\int_{\Omega}r_{\eps}\nabla \phi dx,\ \ \ \forall \phi \in H_0^1(\Omega)\end{equation}
Since $e_{\eps}\in H^1_0(\Omega)$, taking $\phi=e_{\eps}$, we get
\begin{equation}\label{weak form of error equation for 2nd situation}
\int_{\Omega}A^{\eps}\nabla e_{\eps}\nabla e_{\eps} dx=\int_{\Omega}R^{\eps}e_{\eps} dx+\int_{\Omega}r_{\eps}\nabla e_{\eps} dx,\ \ \ \forall \phi \in H_0^1(\Omega)\end{equation}
By the elliptic condition and assumption that $G\in
L^{\infty}(Y)$, the second term in the right hand can be estimated as follows \be
\|r^{\eps}\|^2_{L^2(\Omega)}\leq\eps(\|G\|_{L^{\infty}(Y)}+\Lambda\|N\|_{L^{\infty}(Y)})\|\frac{\partial^2u^0}{\partial
x_j\partial x_k}\|_{L^2(\Omega)}.\ee
By the definition of $F^0(x)$ in (\ref{lemma homomatrix}), we obtain
\begin{eqnarray}&& \nabla_x\cdot( F(x,y)+A(y)\nabla_yw(x,y)-
F^0(x))=\left(\nabla_x\cdot F-\int_Y \nabla_x\cdot
F(x,y)dy\right)\nonumber\\
&& \ \hspace{1cm}+\nabla_x\cdot(A(y)\nabla_yw(x,y))
-\left(\int_Y\nabla_x\cdot( A(y)\nabla_yw(x,y))dy\right)
\end{eqnarray}
It can be easily checked that \begin{equation}\int_Y R^{\eps}(x,y)dy=0.\end{equation}
By lemma \ref{f H-1 estimate} and lemma 1.6 in \cite{Olineak1992}, the first term in the right hand of \eqref{weak form of error equation for 2nd situation} can be estimated as follows:
\begin{equation}
\int_{\Omega}R^{\eps}(x,y)e_{\eps}dx\leq C\eps\|e_{\eps}\|_{H^1_0(\Omega)}
\end{equation}
%

So we obtain
\begin{eqnarray}\label{lemma error control equantion}
\|\nabla(u^{\eps}-u_1-\theta^{\eps})\|_{L^2(\Omega)}&\leq&C\eps\frac
1
{\lambda}(\|G\|_{L^{\infty}}+\Lambda\|N\|_{L^{\infty}})\|\frac{\partial^2u^0}{\partial
x_j\partial x_k}\|_{L^2(\Omega)}\nonumber\\&&+ C\eps
\end{eqnarray}

Next we estimate $\|\nabla \theta^{\eps}\|_{L^2(\Omega)}$.  Multiplying
$\theta^{\eps}+\eps \phi^{\eps}N^j\frac{\partial u^0}{\partial
x_j}+\eps \phi^{\eps}w$ on both sides of \eqref{theta equation}, we
obtain
\begin{eqnarray}\label{lemma theta} \int_{\Omega}|\nabla\theta^{\eps}|^2dx&\leq&
(\frac {\Lambda}{\lambda})^2\int_{\Omega}|\eps
\nabla(N^j\frac{\partial u^0}{\partial x_j}\phi^{\eps})|^2dx+(\frac
{\Lambda}{\lambda})^2\int_{\Omega}|\eps\nabla(w\phi^{\eps})|^2dx\nonumber\\
&=&C(\frac {\Lambda}{\lambda})^2I_1+C(\frac {\Lambda}{\lambda})^2I_2
\end{eqnarray}
where $\phi^{\eps}$ is a cut-off function, satisfying \be \left\{
\begin{array}{l}\label{thetaproperty}
\phi^{\varepsilon}=1,\ \ \mbox{on}\ \partial\Omega\\
0\leq\phi^{\varepsilon}\leq1,\ \ \mbox{on}\  \Omega_{\eps}\\
|\varepsilon\nabla\phi^{\varepsilon}|\leq C\\
\phi^{\varepsilon}\in C^{\infty}(\overline{\Omega}),
\end{array} \right. \ee
with
$ \Omega_{\eps}=\{x\in \Omega \ |\ \mbox{dist}(x,\partial\Omega)\leq\varepsilon\}.$

For $I_1$, we directly use the result in \cite{V.V.Zhikov-2006}
\be\label{lemma I1} I_1\leq C\eps(1+\|N\|_{L^{\infty}(Y)})^2\|\frac
{\partial u^0}{\partial x_j}\|^2_{H^1(\Omega)}.\ee
Please note that  the above estimate only depends on the $\|N(y)\|_{L^\infty(Y)}$  rather than  $\|N(y)\|_{W^{1,\infty}(Y)}$ . This is the contribution of Suslina \cite{T.A.Suslina-2004}.

For $I_2$, we have
\begin{eqnarray}\label{lemma I2}I_2&\leq&\eps^2\int_{\Omega_{\eps}}|\nabla_xw(x,y)|^2dx+
\int_{\Omega_{\eps}}|\nabla_yw(x,y)|^2dx+\int_{\Omega_{\eps}}|w(x,y)|^2dx\nonumber\\
&\leq& C\eps \|w\|^2_ {W^{1,\infty}(\Omega\times  Y)} .
\end{eqnarray}
By (\ref{lemma error control equantion}), (\ref{lemma theta}),
(\ref{lemma I1}), and (\ref{lemma I2}), we complete the proof.\\
\hspace*{\fill}$\Box.$
\begin{remark}\label{1dremark}
Please note that the estimate in (\ref{lemma error control equantion}) is of order $\eps$. The final estimate \eqref{estimate for situation 2} decays to the order $\eps^{\frac 1 2}$, due to the oscillation of the corrector term $\theta^{\eps}$ \eqref{theta equation} at the boundary. For the one-dimension problems, the boundary decays to isolated points and this kind of  oscillation at the boundary does not appear any more. The estimate \eqref{estimate for situation 2} can be improved to
\be \|\partial(u^{\eps}-u_1)\|_{L^2(0,1)}\leq C\eps.\ee
\end{remark}
\begin{remark}
As to the convergence of potential energy, we have as $\eps\rightarrow0 $
\be|\int_{\Omega}\left(\nabla u^{\eps}\cdot A^{\eps}\nabla u^{\eps}+F^{\eps}\cdot\nabla u^{\eps}\right)dx-\int_{\Omega}\left(\nabla u^0\cdot A^0\nabla u^0+F^0\cdot\nabla u^0\right)dx|\rightarrow 0.\ee
\end{remark}
So far we have establish the mathematical homogenization theory for \eqref{physicsmath for situation2}, we now come back to the physics problem \eqref{physics for situation 2}.

By Theorem \ref{lemma on  homgenzition for situation 2 } and unit
transformation,
we get the {\em homogenized} problem of (\ref{physics for situation 2}).
\be \left\{
\begin{array}{l}\label{phyhomo21}
 -\nabla\cdot (K^{0} \nabla p_0(x)) =f^0(x)+\nabla\cdot F^0(x), \ \ \mbox{in}\ \Omega_{D}=(0,D)^{d}, \\
\\
p_0=0, \ \ \ \mbox{on}\ \partial\Omega_{D},
\end{array} \right. \ee
where \be \left\{
\begin{array}{l}
 p_0(x)=D^2\widetilde{p_0}(\hatx)\\
 \\
 K^0=\widetilde{K^0}
\end{array} \right. \label{relationship2} \ee

%
%
 with
$K^0$, $f^0$, and $F^0$
 defined as follows
 \be\label{physics homomatrix 2}\left\{\begin{array}{l}
 K_{ij}^0=\langle K_{ij}(y)+K_{ik}\frac{\partial N^j}{\partial
y_k}\rangle_Y,\\
f^0(x)=\langle f(x, y)\rangle_Y,\\
F^0(x)=\langle F(x,
y)+K(y)\nabla_yw(y)\rangle_Y,\end{array}\right.\ee and $N^j(\frac x
l)=N^j(y)$, $w(x,\frac x l)=w(x,y)$ solving the cell problems
\be\label{lemma cell N}\left\{\begin{array}{l}
-\nabla_y\cdot(K(y)\nabla_y N^j(y))=\nabla_y\cdot(K(y)e^j),\ \mbox{in}\ Y,\\
\\
N^j(y)\ \mbox{is periodic in Y}, \langle N^j\rangle_Y=0
\end{array}\right.\ee
\be\left\{\begin{array}{l}\label{lemma cell w}
-\nabla_y\cdot(K(y)\nabla_y w(x,y))=\nabla_y\cdot(F(x, y)),\ \mbox{in}\ Y,\\
\\
w(x,y)\ \mbox{is periodic in Y}, \langle w\rangle_Y=0.
\end{array}\right.\ee

By unit transformation, \eqref{relationship2}, and  \eqref{estimate for situation 2}, we can obtain the next theorem
\begin{theorem}
If $p$ is the solution of \eqref{physics for situation 2}, $p_1$ is defined as follows  \be p_1=p_0+lN^j\frac {\partial p_0}{\partial x_j}+lw,\ee with $p_0$ solving \eqref{phyhomo21} and $f(x,y)$, $F(x,y)$ are bounded, smooth enough and Y-period with respect to y,
 then there exists a positive number C independent of D, such that
\begin{eqnarray}\label{final physics estimate2 for grad}
\frac 1 {D^3}\int_{\Omega_D}|\nabla(\frac {p} {D}-\frac
{p_1 }{D})|^2dx &\leq& C\frac l
D,
\end{eqnarray}
\begin{eqnarray}\label{l2estimate2}
\frac 1 {D^3} \int_{\Omega_D}|\frac p {D^2}-\frac {p_0}
{D^2}|^2dx&\leq& C(\frac l
D)^2.
\end{eqnarray}
As to the convergence of potential energy, we have
\begin{eqnarray} |E(p)-E_0(p_0)| \rightarrow 0 \ \
\mbox{as}D\rightarrow\infty ,\end{eqnarray} with $${\displaystyle E(p)=\frac 1 {D^3}
\int_{\Omega_D}\left(\nabla \left(\frac p D\right) K\nabla
\left(\frac p D \right)+\frac F D\nabla \left(\frac p
D\right)\right)dx},$$ $$E_0(p_0) = {\displaystyle \frac 1 {D^3} \int_{\Omega_D}\left(\nabla
\left(\frac {p_0} D\right) K^0\nabla \left(\frac {p_0} D
\right)+\frac {F^0} D\nabla \left(\frac {p_0}D\right)\right)dx }.$$
\end{theorem}
Above theorem explains in what sense we can accept the homogenized problem \eqref{phyhomo21}.
\begin{remark}
In one-dimension case, by Remark \ref{1dremark}, our results  are changed to be: \be\label{examp1} \frac 1 D \int_0^D
|\partial_x(\frac p D-\frac {p_0} D)|^2dx\leq C \left(\frac l
D\right)^2,\ee
\be\label{examp0}\frac 1 D \int_0^D |\frac p {D^2}-\frac {p_0}
{D^2})|^2dx\leq C \left(\frac l D\right)^2,\ee
\begin{eqnarray}\label{examenergy} |E(p)-E_0(p_0)| \rightarrow 0 \ \
\mbox{as}\ D\rightarrow\infty ,\end{eqnarray} with $${\displaystyle E(p)=\frac 1 D
\int_0^D\left(\partial_x \left(\frac p D\right) K\partial_x
\left(\frac p D \right)+\frac F D\partial_x \left(\frac p
D\right)\right)dx},$$ $$E_0(p_0) = {\displaystyle \frac 1 D \int_0^D\left(\partial_x
\left(\frac {p_0} D\right) K^0\partial_x \left(\frac {p_0} D
\right)+\frac {F^0} D\partial_x \left(\frac {p_0}D\right)\right)dx }.$$
\end{remark}

\section{Example}
\setcounter{equation}{0}
In this part, we show a one-dimensional example  to verify our
results .
We consider the following problem: \be \left\{
\begin{array}{l}
-\partial _x(K\partial_xp)=-1+\partial_x\left(\left(x+\frac D 2+C\right)K\right)\ \ \mbox{in}\ (0,D),\\
\\
p(0)=p(D)=0,
\end{array} \right. \ee
where ${\displaystyle K=\frac 1 {2+cos(\frac{2\pi x}{l})}}$ and ${\displaystyle C=\frac l
{2\pi}\sin\left(\frac {2\pi D}{l}\right)+\left(\frac l
{2\pi}\right)^2\frac 1 D\cos\left(\frac {2\pi
D}{l}\right)-\left(\frac l {2\pi}\right)^2\frac 1 D}$. It is clear that the source term has micro-structure, which we have discussed in $\S$3.

The  solution is \begin{eqnarray} p&=&\frac{x^2} 2-\frac D 2
x+x\frac l {2\pi}\sin\left(\frac {2\pi x}{l}\right)+\left(\frac l
{2\pi}\right)^2\cos\left(\frac {2\pi
x}{l}\right)\nonumber\\
&&-Cx-\left(\frac {l}{2\pi}\right)^2.\end{eqnarray}

The homogenized problem has the form as \be\left\{\begin{array}{l}
-\partial_x(K^0\partial_x p_0)=-1+\partial_x\left(\frac x 2\right),\ \ \mbox{in}\ (0,D),\\
\\
p_0(0)=p_0(D)=0,
\end{array} \right. \ee
with $K^0=\frac 1 2$ and the   solution  ${\displaystyle p_0=\frac {x^2} 2-\frac D
2 x}$. ${\displaystyle N(\frac x l)=\frac{\sin(\frac {2\pi
x}{l})}{4\pi}}$ is the  solution of cell problem
\be\left\{\begin{array}{l}
-\partial_y(K\partial_y (N+y))=0,\ \ \mbox{in}\ Y=(0,1),\\
\\
N(y)\ \mbox{is Y-periodic with respect to y and}\  \langle
N(y)\rangle_Y=0,
\end{array} \right.\ee
 and ${\displaystyle w(x,\frac x l)=(x+\frac D 2+C)\frac{\sin(\frac
{2\pi x}{l})}{4\pi}}$ is the  solution of cell problem
\be\left\{\begin{array}{l}
-\partial_y(K(y)\partial_yw(x,y))=\partial_y(F(x,y)),\ \mbox{in}\
Y=(0,1),\\
w(x,y)\ \mbox{is Y-periodic with respect to y and}\  \langle
w(x,y)\rangle_Y=0.\end{array}\right.\ee

By direct computation, we obtain
\begin{align}
&\int_0^D(p-p_0)^2dx&\nonumber\\
=&\int_0^D\left(x\frac l {2\pi}\sin\left(\frac
{2\pi x}{l}\right)+\left(\frac l {2\pi}\right)^2\cos\left(\frac
{2\pi
x}{l}\right)\right)^2dx+\int_0^D\left(Cx\right)^2dx
+\int_0^D\left(\frac l {2\pi}\right)^4dx&\nonumber\\
=&D^3\left(\frac
{l^2}{24\pi^2}+\frac{C^2}{3}\right)-D^2\left(\frac{l^3}{16\pi^3}\sin\left(\frac{4\pi
D}{l}\right)\right)+D\left(\frac {l^4} {16\pi^4}-\frac {l^4}{32\pi^4}\cos\left(\frac
{4\pi
D}{l}\right)-\frac{l^2}{8\pi^2}\right)&\nonumber\\
&+\left(\frac {l^5}{128\pi^5}\sin\left(\frac{4\pi
D}{l}\right)-\frac {l^3}{32\pi^3}\sin\left(\frac{4\pi
D}{l}\right)\right).&
\end{align}
So we have \be \frac 1 D\int_0^D(\frac {p-p_0}{D^2})^2dx= \left(\frac l
D\right)^2\left(\left(\frac
{l^2}{24\pi^2}+\frac{C^2}{3}\right)+o(1)\right),\ee
which is consistent with the theoretical result \eqref{examp0}.

If we denote by \begin{eqnarray}\label{example
p1}p_1&=&p_0+lN\partial_x
p_0+lw\nonumber\\
&=&p_0+x\frac l {2\pi}\sin\left(\frac {2\pi x}{l}\right)+
\frac{lC}{4\pi}\sin\left(\frac {2\pi x}{l}\right),\end{eqnarray}
then we have
\begin{eqnarray}
p-p_1&=&\left(\frac l {2\pi}\right)^2\cos\left(\frac {2\pi
x}{l}\right)-\left(\frac l {2\pi}\right)^2-Cx+
\frac{lC}{4\pi}\sin\left(\frac {2\pi x}{l}\right).
\end{eqnarray}
\begin{eqnarray}
\int_0^D(\partial_x(p-p_1))^2dx&=&\int_0^D\left(-\frac l{2\pi}\sin\left(\frac{2\pi x}{l}\right)-C+\frac C 2 \cos\left(\frac{2\pi x}{l}\right)\right)^2dx\nonumber\\
&=&D\left(\frac {l^2}{8\pi^2}+\frac{9C^2}{8}\right)+\left(\frac
{C^2l}{32\pi}-\frac {l^3}{32\pi^2}\right)\sin\left(\frac{4\pi
D}{l}\right).\end{eqnarray} We obtain \be \frac 1
D\int_0^D(\partial_x(\frac{p-p_1} D))^2dx=\left(\frac l
D\right)^2\left(\frac {l^2}{8\pi^2}+\frac{9C^2}{8}+o(1)\right).\ee
This is also consistent with  the theoretical result \eqref{examp1}.

As to the convergence of the  energy, we have
\begin{align}
\ &\int_0^D \partial_x p\cdot K\partial_x pdx+\int_0^D
 F \partial_xpdx&\nonumber\\
 =&\int_0^D fpdx&\nonumber\\
=&\frac {D^3}{12}+\frac{CD^2}{2}+D\left(\frac
{l^2}{4\pi^2}\cos\left(\frac {2\pi D}{l}\right) +\frac
{l^2}{4\pi^2}\right)-\frac l {2\pi}\sin\left(\frac
{2\pi D}{l}\right),&
\end{align}
\begin{eqnarray}
\int_0^D \partial_x p_0\cdot K^0\partial_x p_0dx+\int_0^D
F^0\partial_xp_0dx&=&\frac {D^3}{12},
\end{eqnarray}
so we have \begin{eqnarray} |E(p)-E_0(p_0)| =\frac{C}{D2}+\frac 1{D^2}\left(\frac
{l^2}{4\pi^2}\cos\left(\frac {2\pi D}{l}\right) +\frac
{l^2}{4\pi^2}\right)-\frac l {2\pi D^3}\sin\left(\frac
{2\pi D}{l}\right)\rightarrow 0 \nonumber\end{eqnarray}
as $D\rightarrow\infty$ , with $${\displaystyle E(p)=\frac 1 D
\int_0^D\left(\partial_x \left(\frac p D\right) K\partial_x
\left(\frac p D \right)+\frac F D\partial_x \left(\frac p
D\right)\right)dx},$$ $$E_0(p_0) = {\displaystyle \frac 1 D \int_0^D\left(\partial_x
\left(\frac {p_0} D\right) K^0\partial_x \left(\frac {p_0} D
\right)+\frac {F^0} D\partial_x \left(\frac {p_0}D\right)\right)dx }.$$
This is also consistent with  the theoretical result \eqref{examenergy}.

{\bf Acknowledgment}.   This work  is supported
in part by NSF of China under the Grant 10871190  and by the Qing-Lan Project of Jiangsu Province.

\bibliographystyle{amsplain}

\end{document}